\documentclass[prd,twocolumn,amsmath,amssymb,nofootinbib,floatfix,superscriptaddress]{revtex4}

\usepackage{graphicx,bm,tikz} \usepackage[normalem]{ulem}

\makeatletter
\def\graphicscale{\twocolumn@sw{0.3}{0.4}}
\def\graphicthreescale{\twocolumn@sw{0.3}{0.4}}

\begin{document}

\title{Breaking of the gauge symmetry in lattice gauge theories}

\author{Claudio Bonati} 
\affiliation{Dipartimento di Fisica dell'Universit\`a di Pisa
        and INFN Largo Pontecorvo 3, I-56127 Pisa, Italy}

\author{Andrea Pelissetto}
\affiliation{Dipartimento di Fisica dell'Universit\`a di Roma Sapienza
        and INFN Sezione di Roma I, I-00185 Roma, Italy}

\author{Ettore Vicari} 
\affiliation{Dipartimento di Fisica dell'Universit\`a di Pisa
        and INFN Largo Pontecorvo 3, I-56127 Pisa, Italy}

\date{\today}

\begin{abstract}
We study perturbations that break gauge symmetries in lattice gauge
theories. As a paradigmatic model, we consider the three-dimensional
Abelian-Higgs (AH) model with an $N$-component scalar field and a
noncompact gauge field, which is invariant under U(1) gauge and 
SU($N$) transformations.  We consider gauge-symmetry breaking
perturbations that are quadratic in the gauge field, such as a photon
mass term, and determine their effect on the critical behavior of the
gauge-invariant model, focusing mainly on the continuous transitions
associated with the charged fixed point of the AH field theory.  We
discuss their relevance and compute the (gauge-dependent) exponents
that parametrize the departure from the critical behavior (continuum
limit) of the gauge-invariant model. We also address the critical
behavior of lattice AH models with broken gauge symmetry, showing an
effective enlargement of the global symmetry, from U($N$) to O($2N$),
which reflects a peculiar cyclic renormalization-group flow in the
space of the lattice AH parameters and of the photon mass.
\end{abstract}

\maketitle

{\em Introduction.} Gauge symmetries play a key role in the
construction of the theoretical models of fundamental
interactions~\cite{Weinberg-book,ZJ-book} and in the description of
emergent phenomena in condensed-matter and statistical
physics~\cite{Wen-book,Anderson-15,GASVW-18,Sachdev-19,SSST-19,GSF-19}.
They may be exact, as in the Standard Model of fundamental
interactions, or effectively emerge at low energies, as in some
many-body systems.  Effectively emergent gauge symmetries have also
been discussed in the context of fundamental interactions, see, e.g.,
Refs.~\cite{Wen-book,Wetterich-17,FNN-80,INT-80}. In this case, they
may arise from microscopic interactions of different nature, such as
string models~\cite{Polchinski-book}.

To correctly interpret experimental results in terms of models with an
emergent gauge symmetry, a solid understanding of the effects of
gauge-symmetry violations is essential.  This issue is crucial in the
context of analog quantum simulations, for example, when controllable
atomic systems are engineered to effectively reproduce the dynamics of
gauge-symmetric theoretical models, with the purpose of obtaining
physical information from the experimental study of their quantum
dynamics in laboratory.  Several proposals of artificial
gauge-symmetry realizations have been reported, see, e.g.,
Refs.~\cite{ZCR-15,Banuls-etal-20} and references therein (see also
Refs.~\cite{Martinez-etal-16,Bernien-etal-17,Klco-etal-18,Schweizer-etal-19,
  Gorg-etal-19,Mil-etal-20} for some experimental realizations), in
which the gauge symmetry is expected to effectively emerge in the
low-energy dynamics.  A possible strategy is that of adding a {\em
  penalty} term to the Hamiltonian, that suppresses the interactions
violating the gauge symmetry. This strategy assumes that
gauge-symmetry breaking (GSB) terms become negligible at low energies,
thereby making the dynamics effectively gauge invariant in this
limit~\cite{ZR-11,ZCR-15,BC-20}. In spite of the relevance of these
issues, there is at present little understanding of the effects of
GSB perturbations on the continuum limit of quantum or statistical
systems with gauge symmetries, or equivalently on the critical
behavior close to continuous transitions, where long-range
correlations develop, realizing the corresponding quantum field
theory.

In this paper we address this problem by considering three-dimensional
(3D) lattice gauge theories, obtained by discretizing the action of
corresponding quantum field theories.  We study the role of GSB
perturbations at the critical transitions of gauge-invariant models, to
understand whether and when they are relevant, i.e. they break gauge
invariance in the low-energy or large-distance behavior (continuum
limit). If this is the case, GSB terms may lead to different continuum
limits, as we shall see.

{\em The model.}  As a paradigmatic model, we consider the 3D scalar
electrodynamics or Abelian-Higgs~(AH) field theory, with an
$N$-component complex scalar field $\Phi({\bm x})$ coupled to the
electromagnetic field $A_\mu({\bm x})$. Its Lagrangian density
reads~\cite{ZJ-book}
\begin{equation}
{\cal L} = |D_\mu{\Phi}|^2
+ w \, {\Phi}^*{\Phi} + 
\frac{u}{4} \,({\Phi}^*{\Phi})^2
+ \frac{1}{4 g^2} \, (\partial_\mu A_\nu -
\partial_\nu A_\mu)^2\,,
\label{abhim}
\end{equation}
where $D_\mu \equiv \partial_\mu + i A_\mu$.  The AH theory is
invariant under U(1) gauge and SU($N$) global transformations. Its 3D
renormalization-group (RG) flow has a stable charged (with nonzero
gauge coupling) fixed point (FP) for $N \ge
N_c$~\cite{HLM-74,MZ-03}, with $N_c=7(2)$~\cite{BPV-21,IZMHS-19}.
According to the RG theory~\cite{WK-74,Fisher-75,Wilson-83,PV-02}, the
charged FP is expected to describe the critical behavior, and
therefore the continuum limit, of U(1) gauge models with SU($N$)
global symmetry.

Lattice representations of the continuum theory (\ref{abhim}) differ
for the topological nature of the lattice gauge field.  One can either
use the real field $A_{{\bm x},\mu}$ as in the continuum theory
(noncompact model) or the link variables $\lambda_{{\bm x},\mu}\in
{\rm U}(1)$ (compact model, corresponding to $e^{i A_{{\bm x},\mu}}$).
In this work we mostly consider the 3D noncompact AH (ncAH) model
defined on cubic lattices of size $L^3$, which has a continuous
transition line for $N>N_c$, along which the continuum limit is
described by the 3D AH field theory
(\ref{abhim})~\cite{BPV-21,PV-19-AH3d,footnote-highercha}.  The
fundamental fields are unit-length $N$-component complex vectors ${\bm
  z}_{\bm x}$ ($\bar{\bm z}_{\bm x}\cdot {\bm z}_{\bm x}=1$) defined
on the lattice sites ${\bm x}$ and real fields $A_{{\bm x},\mu}$
defined on the lattice links.  The lattice action is
\begin{eqnarray}
S_{\rm AH}({\bm z},{\bm A}) &=&
- J \, N
\sum_{{\bm x},\mu} 2\,{\rm Re}\,( \bar{\bm z}_{\bm x} \cdot 
\lambda_{{\bm x},\mu}\,{\bm z}_{{\bm x}+\hat\mu})  
\label{ncah}\\ 
&&\;+ {1\over 4 g_0^2} \sum_{{\bm x},\mu\nu} (\Delta_{\mu} A_{{\bm
    x},\nu} - \Delta_{\nu} A_{{\bm x},\mu})^2\,, \nonumber
\end{eqnarray}
where $\lambda_{{\bm x},\mu} \equiv e^{iA_{{\bm x},\mu}}$, $g_0$ is
the lattice gauge coupling, $\hat{\mu}$ are unit vectors along the
lattice directions, and $\Delta_\mu A_{{\bm x},\nu} = A_{{\bm
    x}+\hat{\mu},\nu}- A_{{\bm x},\nu}$.  The action $S_{\rm AH}$ has
a global SU($N$) symmetry, ${\bm z}_{\bm x} \to V {\bm z}_{\bm x}$
with $V\in\mathrm{SU}(N)$, and a local U(1) gauge symmetry, ${\bm
  z}_{\bm x} \to e^{i\theta_{\bm x}} {\bm z}_{\bm x}$ and $A_{{\bm
    x},\mu}\to A_{{\bm x},\mu} + \theta_{\bm x}-\theta_{{\bm
    x}+\hat{\mu}}$.  We consider $C^*$ boundary
conditions~\cite{BPV-21,KW-91,LPRT-16} (see also App.~\ref{Supplmat}),
to remove the degeneracy under $A_{{\bm x},\mu}\to A_{{\bm x},\mu} +
2\pi n_{\mu}$ with $n_{\mu}\in\mathbb{Z}$, obtaining well defined
expectation values for gauge-invariant operators $O({\bm z},{\bm A})$,
\begin{equation}
  \langle O({\bm z},{\bm A}) \rangle = { \sum_{\{{\bm z},{\bm A}\}}
    O({\bm z},{\bm A})
    \;
    e^{- S_{\rm AH}({\bm z},{\bm A})}\over
  \sum_{\{{\bm z},{\bm A}\}} e^{- S_{\rm AH}({\bm z},{\bm A})}}\,.
\label{zah}
\end{equation}
The phase diagram of the ncAH model (\ref{ncah}) with $N\ge 2$ is
characterized by a Coulomb phase for small $J$ (short-ranged scalar
and long-ranged gauge correlations), a Higgs phase for large $J$ and
small $g_0$ (condensed scalar-field and gapped gauge correlations),
and a molecular phase for large $J$ and $g_0$ (condensed scalar-field
and long-ranged gauge correlations)~\cite{BPV-21}.  They are separated
by three transition lines, which are continuous or of first order
depending on $N$.  In particular, for $N>N_c=7(2)$, the ncAH model
undergoes continuous transitions between the Coulomb and Higgs (CH)
phases, for $0<g_0^2\lesssim 4$. The corresponding critical behavior
is described by the charged FP of the 3D AH field
theory~\cite{BPV-21}.  For $g_0\to 0$, one has $A_{{\bm x},\mu} \to 1$
modulo gauge transformations, so that one recovers the O($2N$) vector
model.  We consider the gauge-invariant bilinear operator
\begin{equation}
Q_{{\bm x}}^{ab} = \bar{z}_{\bm x}^a z_{\bm x}^b - {1\over N}
\delta^{ab}\, ,
\label{qdef}
\end{equation}
which transforms as $Q_{{\bm x}} \to {V}^\dagger Q_{{\bm x}} \,{V}$
under global SU($N$) transformations. It provides an effective order
parameter for the spontaneous breaking of the global SU($N$) symmetry.

{\em GSB perturbations.}  We study how perturbations breaking the U(1)
gauge symmetry affect the CH critical behavior.  In this exploratory
study we consider the quadratic perturbation
\begin{eqnarray}
  P_{M} = {r \over 2} \sum_{{\bm x},\mu} A_{{\bm x},\mu}^2\,,
\label{defmA}
\end{eqnarray}
which can be interpreted as a photon mass term. Such a mass term is
generally introduced as an infrared regulator in perturbative
computations in quantum electrodynamics~\cite{ZJ-book}. We also
consider the local quadratic operators
\begin{eqnarray}
P_{L}= {a \over 2} \sum_{{\bm
    x}}\,(\sum_\mu \Delta_\mu A_{{\bm x},\mu})^2,\;\;
P_{A} = {b \over 2} \sum_{{\bm x}}\,(\sum_\mu n_\mu \,A_{{\bm x},\mu})^2,
\;\;
\label{defGA}
\end{eqnarray}
where $n_\mu$ is an arbitrary unit vector.  When added to the ncAH
action, i.e., if one considers $S=S_{\rm AH}+P_\#$, all quadratic
terms defined in Eqs.~(\ref{defmA}) and (\ref{defGA}) break gauge
invariance, leaving a global U($N$) symmetry ${\bm z}_{\bm x} \to U
{\bm z}_{\bm x}$, $U\in\mathrm{U}(N)$. However, they affect the
critical behavior quite differently. The mass term
  (\ref{defmA}) is expected to be relevant at the CH transitions,
  since it drastically changes the long-distance properties of the
  gauge-field correlations. In particular, the Coulomb phase
  disappears in the presence of a photon mass.  Therefore, as soon as
  the perturbation is turned on ($r > 0$), the system is expected to
  flow out of the charged AH FP. On the other hand, the quadratic
terms $P_L$ and $P_A$, cf. Eq.~(\ref{defGA}), may be interpreted as
the result of the Fadeev-Popov procedure for a gauge
fixing~\cite{ZJ-book}, being related to the Lorentz ($\partial_\mu
A_\mu=0$) and axial (${\bm n}\cdot {\bm A}=0$) gauge
fixing~\cite{footnote-gf}, respectively.  If they are the only GSB
perturbations present in the model, they are expected to be irrelevant
for gauge-invariant correlations (more precisely, their presence does
not change gauge-invariant expectation values).  However, as we shall
see below, they play a role, when they are added to the action
together with the mass term (\ref{defmA}), as they make the limit
$r\to 0$ well defined.

{\em Relevance of the GSB perturbations.}  To characterize the
strength of the perturbation $P_M$, we compute the corresponding RG
dimension $y_r>0$. This exponent provides information on how to scale
$r$ to keep GSB effects small. Indeed, when the correlation length
$\xi$ increases, approaching the continuum limit, one should decrease
$r$ faster than $\xi^{-y_r}$ to ensure that GSB effects are negligible.  
We estimate $y_r$ by finite-size scaling (FSS) analyses of Monte
Carlo (MC) data.  We consider the correlation function $\langle {\rm
  Tr}\, Q_{\bm x}Q_{\bm y} \rangle$ of the operator $Q_{\bm x}$
defined in Eq.~(\ref{qdef}), and the corresponding second-moment
correlation length $\xi$.  We consider RG-invariant quantities $R$,
such as $R_\xi=\xi/L$ and the Binder parameter $U = {\langle
  \mu_2^2\rangle}/{\langle \mu_2 \rangle^2}$, where $\mu_2 =
\sum_{{\bm x},{\bm y}} {\rm Tr}\, Q_{{\bm x}} Q_{\bm y}$.  At
continuous transitions driven by the parameter $J$, they are expected
to behave as~\cite{PV-02}
\begin{equation}
R(L,J,g_0) \approx f_R(X)+O(L^{-\omega})\,,\;\; X =(J-J_c)\,L^{1/\nu}\,,
\label{rljg0}
\end{equation}
where $\nu$ is the length-scale critical exponent, and $\omega>0$ is
the exponent controlling the leading scaling corrections.  It is also
useful to consider the FSS relation~\cite{BPV-19-sqcd}
\begin{equation}
  U =F_U(R_\xi) + O(L^{-\omega})\, ,
  \label{uvrxi}
\end{equation} 
where $F_U$ is a universal function independent of any normalization.
To estimate $y_r$, we consider the behavior of the RG invariant
quantities $R$ in the presence of the GSB term (\ref{defmA}). In 
the large-$L$ limit, we expect  \cite{Fisher-MCP}
\begin{equation}
R(L,J,g_0,r) \approx {\cal F}_R(X,Y)\,,\qquad Y=r L^{y_r}\,,
\label{tfrxy}
\end{equation}
which holds provided \cite{MCP} that $y_r > 1/\nu$, where $\nu$ is the
thermal exponent of the gauge model (along the CH transition line we
have $1/\nu =1.387(6)$, 1.247(12) for $N=15$, 25, respectively).
Eq.~(\ref{tfrxy}) is the usual FSS relation for a multicritical point
in systems with a global symmetry.  However, in the present case its
validity is not obvious, given that the mass term $P_M$ is not well
defined in the ($r=0$) gauge-invariant noncompact theory: averages of
the mass term can only be computed in the presence of a maximal gauge
fixing \cite{Creutz-book,footnote-gf}, such as the axial (using $C^*$
conditions) or Lorentz ones.  For these reasons, we consider three
  different actions with GSB terms:
\begin{eqnarray}
  &{\rm M}1: \quad &S_1 = S_{\rm AH} + P_M\,,\label{m1def}\\
  &{\rm M}2: \quad &S_2 = S_{\rm AH} + P_L + P_M\,, \label{m2def}\\
  &{\rm M}3:  \quad &S_3 = S_{\rm AH} + P_M\;\;\;
  {\rm with}\;\;A_{{\bm x},3} = 0\,,
  \quad \label{m3def}
\end{eqnarray} 
where M2 can be associated with the Lorentz gauge, and M3 is defined
imposing the axial gauge.  We expect Eq.~(\ref{tfrxy}) to be well
defined in models M2 and M3, while its validity in model M1 is instead
not clear.

\begin{figure}[tbp]
\includegraphics*[width=0.8\columnwidth]{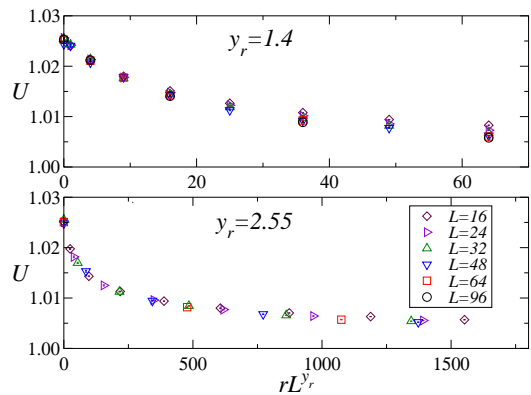}
\caption{Data of $U$ at the critical point $J_c\approx 0.295515$ of
  the ncAH model for $N=25$ and $g_0^2=2.5$, as a function of $Y=r
  L^{y_r}$. Results for models M1 (without gauge fixing, top) and M3
  (axial gauge, bottom). }
\label{figyr}
\end{figure}

{\em Numerical estimates of the RG dimensions.}  We performed MC
simulations for $N=15$ and $N=25$ along the CH transition line
(estimates of the critical points and exponents can be found in
Ref.~\cite{BPV-21}), for the three models M\#, see App.~\ref{Supplmat} for
details.  The results confirm that $P_M$ is relevant. Indeed, for
fixed $r$, there is a clear departure from the gauge-invariant ($r=0$)
critical behavior. In Fig.~\ref{figyr} we show results for $N=25$ at
the critical point.  The exponent $y_r$ is estimated by fitting the
data at $J_c$ to Eq.~(\ref{tfrxy}), setting $X=0$. We obtain $y_r =
1.4(1)$ for M2 [for both $a=1$ and $a=10$, cf. Eq.~(\ref{defGA})] and
$y_r = 2.55(5)$ for M3.  We also mention that if we apply
Eq.~(\ref{tfrxy}) to $U$ computed in M1 without gauge fixing, we
obtain the effective estimate $y_r\approx 1.4$, see top
Fig.~\ref{figyr}, confirming the relevance of $P_M$ along the CH
transition line.  Analogous results are obtained for $N=15$, in
particular $y_r=2.55(10)$ for M3. The exponent $y_r$ turns out to
depend on the gauge fixing, indicating that the gauge fixing
influences the RG properties of the mass perturbation. Apparently,
gauge-dependent modes, that are controlled by the gauge fixing term, 
are crucial in determining the effects of the photon mass term. 
Note that $y_r$ is quite large, therefore the
corresponding GSB perturbation must decrease rapidly with $L$
---faster than $L^{-y_r}$---to keep GSB effects under control.

{\em Critical behaviors in the presence of finite GSB terms.}  We
now address the behavior of the ncAH model in the presence of a finite
GSB term such as the photon mass. Also for finite $r$ we expect a
transition at a finite value $J_c(r)$, with $J_c(r=0) = J_c$, where
$J_c$ is the CH transition point in the gauge-invariant model.  Since
the charged fixed point is unstable with respect to $P_M$, we expect
the transition to belong to a different universality class, which
should only depend on the global symmetry of the model.  Although the
global symmetry group for $r > 0$ is U($N$), we will now argue that
continuous transitions at $J_c(r)$ are characterized by a larger
O($2N$) invariance group.  We note that, since gauge fields are not
expected to be relevant for $r\not=0$, one can use the standard
Landau-Ginzburg-Wilson (LGW)
approach~\cite{WK-74,Fisher-75,Wilson-83,PV-02} to predict the
critical behavior. Since the gauge symmetry is broken, ${\bm z}_{\bm
  x}$ represents the microscopic order-parameter field. Therefore, the
LGW basic field is an $N$-component complex vector $\Psi({\bm
  x})$. The Lagrangian is the sum of the kinetic term
$|\partial_\mu\Psi|^2$ and of the most general U($N$)-invariant
quartic potential:
\begin{eqnarray}
{\cal L}_{\rm LGW} =  \partial_\mu \Psi^*\cdot\partial_\mu \Psi
 + w \;\Psi^*\cdot\Psi   + {u\over 4} \,(\Psi^*\cdot\Psi)^2\,.
\label{UNLGW}
\end{eqnarray}
It is easy to check that ${\cal L}_{\rm LGW}$ is actually O($2N$)
invariant. Indeed, there are no dimension-2 and 4 U($N$) invariant
operators that break the O($2N$) symmetry. The lowest-dimension
operators that are not O($2N$) symmetric have dimension six close to
four dimensions---for instance,
$(\mathrm{Im}\,\Phi^*\cdot\partial_{\mu}\Phi)^2$---and thus they are
expected to be irrelevant at the 3D O($2N$) FP.  Therefore, the
critical behavior of generic vector systems with global U($N$)
invariance (without gauge symmetries) is expected to belong to the
O($2N$) universality class, implying an effective enlargement of the
global symmetry of the critical modes (restricted only to the critical
region).

The above analysis can be extended to lattice AH models with compact
gauge fields (cAH), using the link variables $\lambda_{{\bm x},\mu}\in
{\rm U}(1)$ and the pure gauge action $S_{\lambda} = - g_0^{-2}
\sum_{{\bm x},\mu\neq\nu} {\rm Re}\,\lambda_{{\bm x},{\mu}}
\,\lambda_{{\bm x}+\hat{\mu},{\nu}} \,\bar{\lambda}_{{\bm
    x}+\hat{\nu},{\mu}} \,\bar{\lambda}_{{\bm x},{\nu}}$ in
Eq.~(\ref{ncah}). Unlike ncAH models, cAH models with $N\ge 2$ present
only two phases, separated by a disorder-order transition line where
gauge correlations are not critical~\cite{PV-19-AH3d}.  Since the
scalar fields turns out to be the only critical degrees of freedom,
the effective description of the transitions is provided by the
SU($N$)-invariant LGW $\Phi^4$ theory with a matrix gauge-invariant
order parameter, corresponding to $Q_{\bm x}^{ab}$ in
Eq.~(\ref{qdef})~\cite{PV-19-CP,PV-19-AH3d}.  For $N=2$ this LGW
theory has a stable O(3) vector FP, thus predicting O(3) continuous
transitions~\cite{PV-02} for any gauge coupling $g_0>0$, including
$g_0\to\infty$ [for $g_0\to 0$, instead, the model becomes equivalent
  to the O(4) vector model]. This has been also confirmed
numerically~\cite{PV-19-AH3d}.  Gauge invariance can be broken by
adding $P_M=- r \sum_{{\bm x},\mu} {\rm Re} \,\lambda_{{\bm x},\mu}$,
which plays the role of a photon mass for $\lambda_{{\bm x},\mu}$
close to 1. When the gauge symmetry is effectively broken 
(as discussed in Ref.~\cite{BPV-inprep}, this requires $r$ to be 
sufficiently large), the
critical behavior should be described by the LGW theory (\ref{UNLGW}),
which predicts that continuous transitions belong to the O(4) vector
universality class.

\begin{figure}[tbp]
\includegraphics*[width=0.8\columnwidth]{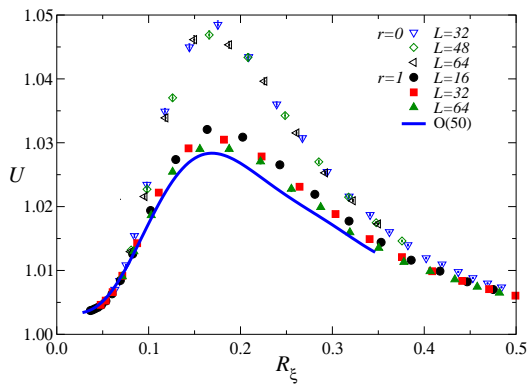}
\caption{Estimates of $U$ versus $R_\xi$ for the model M1 with $N=25$,
  $g_0^2=2.5$, $r =1$.  We also report results for the gauge-invariant
  model ($r=0$) and the O(50) vector model (full line, obtained by
  large-$L$ extrapolations of MC data for the appropriate spin-2
  correlations, see Ref.~\cite{PV-19-AH3d} and App.~\ref{Supplmat}).  The
  results for $r=1$ appear to converge toward the O(50) universal
  curve, consistently with $O(L^{-\omega})$ corrections with
  $\omega\approx 1$, supporting the RG prediction reported in the
  text.}
\label{r1crit}
\end{figure}

The RG predictions at fixed $r$ are confirmed by numerical results for
both ncAH and cAH models.  In Fig.~\ref{r1crit} we plot $U$ versus
$R_\xi$ for the ncAH model with $N=25$ and $r=1$. The data around the
critical point $J_c(r)$ are expected to converge to a universal curve,
cf. Eq.~(\ref{uvrxi}), which can be compared with the analogous curves
of models that belong to known universality classes.  The data
approach the O($2N$) vector universal curve (obtained using an
appropriate operator that corresponds to $Q_{\bm x}^{ab}$ in the
O($2N$) model~\cite{PV-19-AH3d}), confirming the LGW RG argument. For
the cAH model with $N=2$, we observe an asymptotic O(4) vector
behavior for $r = 1$ and $r = 2.25$, in agreement with the general
arguments in App.~\ref{Supplmat}. 

{\em Various classes of GSB perturbations.}  On the basis of the
results presented in this paper, we may distinguish three classes of
GSB perturbations.  (i)~First, there are GSB perturbations that are
relevant at the stable FP of the lattice gauge-invariant theory. They
drive the system out of criticality and may give rise to a different
critical behavior. The photon mass term (\ref{defmA}) plays this role
along the CH line in the ncAH model.  (ii)~A second class corresponds
to gauge fixings and GSB perturbations like those appearing in
Eq.~(\ref{defGA}).  If they are the only GSB terms present in the
model, they are irrelevant: gauge-invariant observables are
unchanged. However, if they are present together with some relevant
GSB perturbation, they play a role: the RG flow close to the charged
FP depends both on the gauge-fixing and on the relevant
perturbation. This may be due to the fact that a gauge fixing is
needed to make non-gauge-invariant correlations well defined in the
gauge-invariant theory or to the role of gauge-dependent modes that are 
sensitive to gauge fixings.  (iii)~GSB perturbations associated with RG
operators with negative RG dimensions, whose effects are suppressed in
the critical (continuum) limit.

When the added GSB perturbations are relevant, the lattice system may
develop a different critical behavior or continuum limit. This is the
case of the ncAH model with a photon mass term, which has a global
U($N$) invariance.  Quite interestingly, the transitions in this model
belong to the O($2N$) vector universality class, with an effective
enlargement of the global symmetry at the transition.  This symmetry
enlargement is expected in any model in which the GSB perturbation is
relevant and it preserves the global U($N$) symmetry.

\begin{figure}[tbp]
\includegraphics*[width=0.8\columnwidth]{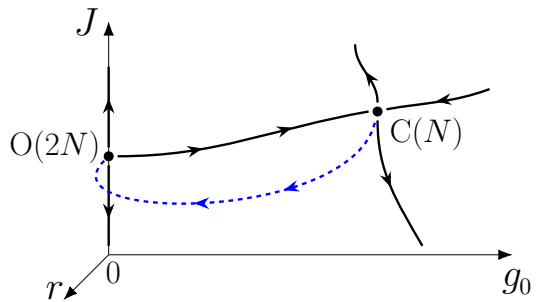}
\caption{Sketch of the cyclic RG flow of the ncAH model in the space
  of the parameters $J$, $g_0$ and $r$, showing an unusual loop
  between the O($2N$) and the charged FP C($N$).  }
\label{rgsketch}
\end{figure}

{\em Cyclic RG flow.}  It is worth noting that the above results lead
to a peculiar RG flow, see Fig.~\ref{rgsketch} for a sketch in the
coupling space $(J,g_0,r)$.  For $g_0^2\to 0$, the gauge fields are
frozen, and the model is equivalent to the O($2N$) vector model, whose
critical behavior is controlled by the corresponding O($2N$) FP. If
the gauge interactions are turned on, i.e., one sets $g_0>0$ keeping
$r=0$, the systems flows towards the charged FP of the AH field
theory, which is stable for any $0<g_0^2\lesssim 4$.  Finally, if a
photon mass is added, i.e., one sets $r > 0$, since the charged FP is
unstable under this perturbation, the RG flow goes back to the O($2N$)
FP, which is now stable, independently of $g_0$ and $r$. This RG
behavior can be hardly reconciled with an irreversibility of the RG
flow, analogous to that generally associated with the monotonic
properties implied by the $c$-theorem of 2D critical
systems~\cite{Zamolodchikov-1986,Cardy-book}, see also
Refs.~\cite{Pufu-17,Grover-14,CH-12,KPSS-12,KPS-11,MS-11} for
similar proposals in 3D systems~\cite{footnoteirrflow}.

{\em Conclusions.}  In conclusion, we have studied the effect of GSB
perturbations on the critical behavior---or, equivalently, the
continuum limit---of gauge-invariant theories. The behavior at charged FP 
turns out to be more complicated than that observed when global symmetries are
broken. In particular, we observe apparent violations of
universality. For instance, the RG dimension of the same GSB
perturbation appears to depend on local gauge-fixing conditions, a
result that, we believe, should be further investigated.  Moreover,
GSB perturbations give rise to unexpected phenomena, like the cyclic
RG flow sketched in Fig.~\ref{rgsketch}.

Several extensions are called for, to achieve a satisfactory
understanding of the problem and to identify its universal features,
such as the study of other lattice gauge theories---in particular, it
would be interesting to extend the analysis to the nonabelian gauge
groups--- and of other classes of GSB perturbations, for example
preserving residual discrete gauge subgroups (such approximations may
be useful for analog simulations).  It would also be important to
rephrase and extend the present results to quantum Hamiltonian systems
\cite{Kogut-79} (see Refs.~\cite{VHH-20,ZVHHB-21} for recent works
addressing issues related to GSB effects and the approach to the
continuum limit).

\bigskip

\acknowledgments

Numerical simulations have been performed on
the CSN4 cluster of the Scientific Computing Center at INFN-PISA.

\appendix

\section{The numerical analyses}
\label{Supplmat}

In this appendix we provide some details on the numerical computations
reported in the paper.

\subsection{The models}

In most of the simulations we have considered the noncompact
Abelian-Higgs (ncAH) model.  The fundamental fields are unit-length
$N$-component complex vectors ${\bm z}_{\bm x}$ ($\bar{\bm z}_{\bm
  x}\cdot {\bm z}_{\bm x}=1$) defined on the lattice sites and real
fields $A_{{\bm x},\mu}$ defined on the lattice links.  The lattice
action is
\begin{equation}
S_{\rm ncAH}({\bm z},{\bm A}) = S_{z}({\bm z},{\bm A}) + S_{nc}({\bm
  A}) \,,
\end{equation}
where 
\begin{eqnarray}
S_z({\bm z},{\bm A}) &=&- J \, N
\sum_{{\bm x},\mu} 2\,{\rm Re}\,( \bar{\bm z}_{\bm x} \cdot 
\lambda_{{\bm x},\mu}\,{\bm z}_{{\bm x}+\hat\mu})   \,,
\label{suppl-ncah} \\ 
S_{nc}({\bm A}) &=& {1\over 4 g_0^2} \sum_{{\bm x},\mu\nu} (\Delta_{\mu} A_{{\bm
    x},\nu} - \Delta_{\nu} A_{{\bm x},\mu})^2\,, \nonumber
\end{eqnarray}
$\lambda_{{\bm x},\mu} \equiv e^{iA_{{\bm x},\mu}}$, $g_0$ is the
lattice gauge coupling, $\hat{\mu}$ are unit vectors along the lattice
directions, and $\Delta_\mu A_{{\bm x},\nu} = A_{{\bm
    x}+\hat{\mu},\nu}- A_{{\bm x},\nu}$.

We have also considered the lattice compact AH model (cAH), with action
\begin{equation}
  S_{\rm cAH}({\bm z},{\bm \lambda}) = S_{z}({\bm z},{\bm \lambda})
  +  S_{c}({\bm \lambda})\,,
\end{equation}
where $S_{z}({\bm z},{\bm \lambda})$ is given by
Eq.~(\ref{suppl-ncah}), while
\begin{equation}
S_{c}({\bm \lambda}) = -
g_0^{-2} \sum_{{\bm x},\mu\neq\nu} {\rm Re}\,\lambda_{{\bm x},{\mu}}
\,\lambda_{{\bm x}+\hat{\mu},{\nu}} \,\bar{\lambda}_{{\bm
    x}+\hat{\nu},{\mu}} \,\bar{\lambda}_{{\bm x},{\nu}} \,.
\end{equation}
To simulate the ncAH model, it is not possible to use periodic
boundary conditions, since all gauge-invariant observables associated
to loops that wrap around the lattice are not bounded and their
average values are ill-defined. As in our previous work~\cite{BPV-21},
we consider $C^*$ boundary conditions. They are used here for the cAH
model too, although periodic boundary conditions would be appropriate,
as well.  We consider cubic lattices of size $L$, so that $C^*$
boundary conditions amount to the identifications (see
Ref.~\cite{BPV-21} for a thorough discussion)
\begin{equation} 
\label{suppl-cstardef}
A_{{\bm x} + L \hat{\nu}, \mu} = - A_{{\bm x}, \mu}\ , \qquad {\bm
  z}_{{\bm x} + L \hat{\nu}} = \bar{\bm z}_{\bm x} \, .
\end{equation}
To be consistent with Eq.~\eqref{suppl-cstardef}, local gauge
transformations are defined by $A_{{\bm x}, \mu} \to A_{{\bm x}, \mu}
+ \alpha({\bm x} + \hat{\mu}) - \alpha({\bm x})$, with an antiperiodic
function $\alpha({\bm x})$: $\alpha({\bm x} + L \hat{\nu}) = -
\alpha({\bm x})$.  As a consequence, observables that involve a
nontrivial wrapping around the lattice are not gauge invariant.

$C^*$ boundary conditions are very convenient when implementing axial
gauges.  Indeed, it is possible to fix $A_{{\bm x},3}=0$---or, more
generally, $\sum_\mu n_\mu A_{{\bm x},\mu}=0$--- on all lattice sites,
at variance with the case of periodic boundary conditions.  From the
explicit construction discussed in Ref.~\cite{BPV-21}, it follows that
no residual gauge freedom is left once $A_{{\bm x},3}=0$ is enforced
on all lattice sites.  It is important to note that also the Lorentz
gauge is a maximal gauge, with no residual gauge freedom. Indeed,
suppose the opposite, i.e., that there are two different gauge
configurations $A_{{\bm x},\mu}^{(1)}$ and $A_{{\bm x},\mu}^{(2)}$
that are related by a gauge transformation, $A_{{\bm x}, \mu}^{(2)} =
A_{{\bm x}, \mu}^{(1)} + \Delta_\mu \alpha({\bm x})$, and that both
satisfy the condition $\sum_\mu \Delta_\mu A_{{\bm x},\mu}^{(i)}=0$.
The function $\alpha({\bm x})$ must satisfy
\begin{equation}
   \sum_\mu \Delta_\mu [\Delta_\mu \alpha({\bm x})] = 0\,,
\end{equation}
which implies that $\alpha$ is a zero eigenmode of the lattice
Laplacian. For $C^*$ boundary conditions there is no zero mode, as
$\alpha$ is antiperiodic, proving that $A_{{\bm x},\mu}^{(1)} =
A_{{\bm x},\mu}^{(2)}$.  For periodic boundary conditions, there is
one zero mode, $\alpha({\bm x}) = C$, where $C$ is space independent,
so that also in this case $A_{{\bm x},\mu}^{(1)} = A_{{\bm
    x},\mu}^{(2)}$.

We have considered the ncAH model for $N=15$ and $25$, fixing in both cases
$g_0^2 = 2.5$. For this value of $g_0$ the ncAH model undergoes a continuous
transition for $J = J_c$. As discussed in Ref.~\cite{BPV-21}, such transition
is controlled by the charged fixed point (FP) of the Abelian-Higgs field
theory.  We have performed simulations for $J=J_c$ on lattices of size $L\le
96$ ($N=25$) and $L\le 64$ ($N=15$). For $N=15$ we used the estimate $J_c$
reported in Ref.~\cite{BPV-21}, see Table~\ref{suppl-tab:critcoup}.  For $N=25$
we used an improved estimate.  We performed additional simulations for
$J\approx J_c$ on larger lattices (while in Ref.~\cite{BPV-21} we limited
ourselves to lattices with $L\le 64$, here we consider values of $L$ up to
$L=96$) and reanalyzed the data. The result is $J_c = 0.295515(4)$, which is
consistent with the estimate $J_c = 0.295511(4)$ reported in
Ref.~\cite{BPV-21}.  We have also considered the cAH model for $N=2$, the only
case where a continuous transition is present. The parameter $g_0^2$ does not
play any role \cite{PV-19-AH3d} and we have therefore set $1/g_0^2 = 0$ (no
gauge action); the estimate of the corresponding critical value $J_c$ is also
reported in Table~\ref{suppl-tab:critcoup}.

Beside simulations of the gauge model, we have also performed
simulations of the O($2N$) spin model with action $S_z$ and
$\lambda_{{\bm x},\mu} = 1$, measuring the same quantities we compute
in the gauge model (see Appendix B of Ref.~\cite{PV-19-AH3d} for a
discussion of the relation between correlation functions of the
CP$^{N-1}$ order parameter $Q^{ab}$ and spin-two correlations in the
vector O($2N$) spin model). We considered $N=2$ and $N=25$,
determining $U$ and $R_\xi$, and in particular the universal curve $U
= F(R_\xi)$.

\begin{table}[t]
\begin{tabular}{rcl}
\hline\hline 
$N$\hspace{0.5cm} & $g_0^2$\hspace{0.5cm} & \multicolumn{1}{c}{$J_c$}  \\
\hline
25  & 2.5 & 0.295515(4) \\
15  & 2.5 & 0.309798(6) \\
2   & $\infty$ & 0.7102(1)   \\ 
\hline\hline
\end{tabular}
\caption{Critical values $J_c$ of the coupling $J$ for the values of
  $g_0^2$ used in the present simulations. The value of $J_c$ for
  $N=25$ is an improvement of the estimate of Ref.~\cite{BPV-21}.
  Results for $N=15$ and $N=2$ are taken from Ref.~\cite{BPV-21} (note
  that $1/g_0^2$ was named $\kappa$), and Ref.~\cite{PV-19-CP},
  respectively.  }
\label{suppl-tab:critcoup}
\end{table}

\begin{figure}[btp]
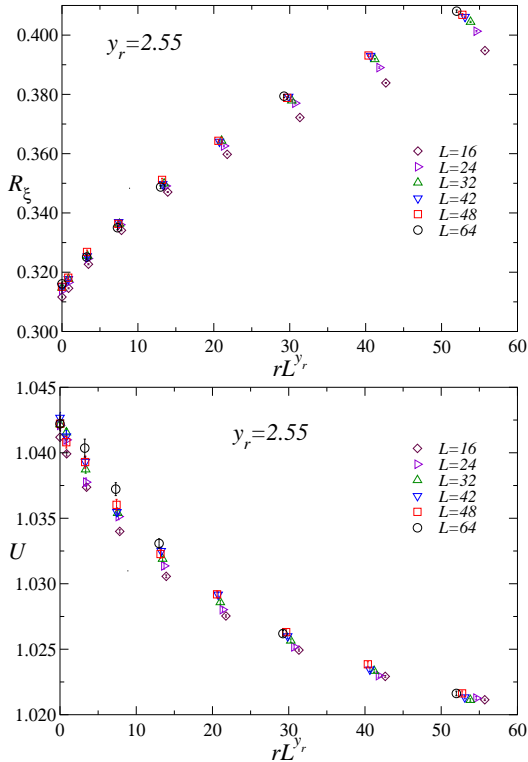

\includegraphics*[width=0.8\columnwidth]{rxim_nccp14_kappa0p4_J0p309798.eps}
\includegraphics*[width=0.8\columnwidth]{um_nccp14_kappa0p4_J0p309798.eps}
\caption{Estimates of $R_\xi$ (top) and $U$ (bottom) at the critical
  point $J_c\approx 0.309798$ of the ncAH model for $N=15$ and
  $g_0^2=2.5$, as a function of $r L^{y_r}$, for model M3 (axial
  gauge). }
\label{suppl-figyrN15}
\end{figure}

\subsection{Technical details: simulations and data analysis}

In the Monte Carlo simulations we use an overrelaxation algorithm, obtained by
combining Metropolis updates of the scalar and of the gauge fields and
microcanonical updates of the scalar field. The latter moves are obtained by
generalizing the usual reflection moves used in O($N$) models. We perform a
Metropolis update of the ${\bm z}$ and gauge fields every 5 (on the largest
lattices, every 10) microcanonical updates of the scalar field. Trial states
for the Metropolis updates are generated by adding a random number to $A_{{\bm
x},\mu}$ and by multiplying ${\bm z}_{\bm x}$ by a random $2\times 2$ unitary
matrix close to the identity; in both cases, we tune the update to have an
acceptance rate of approximately $30\%$.

The typical statistics (discarding thermalization) of each data point
varies in the range $4\times 10^5$-$2\times 10^6$.  Errors are
estimated by using a combination of jackknife and blocking procedures.
In all cases, the autocorrelation times were at most of the order of
$10^2$ updates.

To compute the estimates of $y_r$, we have assumed that $U$ and $R_\xi$ satisfy
the scaling relation
\begin{equation}
   R(r,L) = f(r L^{y_r}) + L^{-\omega} g(r L^{y_r})
\label{suppl-FSSA}
\end{equation}
at the critical point $J = J_c$. In the fits we have approximated the
scaling functions with polynomials and we have analyzed $U$ and
$R_\xi$ together, looking for a value of $y_r$ the minimizes the sum
of the residuals for the two observables.  The value of $\omega$ is
unknown. In the absence of GSB terms, we expect $\omega$ to be close
to 1 (for $N=\infty$ we have $\omega =1$), but, once GSB terms are
added, new irrelevant operators may appear and $\omega$ may be
smaller. For this reason we have determined $y_r$ for values of
$\omega$ in the interval 0.2-1. Except for model M1, the $\omega$
dependence is at most of the same order of the statistical errors.

\subsection{Results for $N=15$}

To investigate the $N$ dependence of the exponent $y_r$, beside
considering the ncAH model with $N=25$ we have also considered the
same model for $N=15$.  We have only studied model M3 (axial gauge),
obtaining
\begin{equation}
y_r=2.55(10)\,,
\end{equation}
which should be compared with the result for $N=25$, $y_r=2.55(5)$.
The $N$ dependence is apparently small, much smaller than the
statistical errors.  In Fig.~\ref{suppl-figyrN15} we show $R_\xi$ and
$U$ against $rL^{y_r}$ (the analogous plot of $U$ for $N=25$ is shown
in the main text).  The ratio $R_\xi$ shows a very nice scaling
behavior with small scaling corrections, which increase as $rL^{y_r}$
increases. The Binder parameter shows larger corrections, that have
the opposite behavior: they decrease as $rL^{y_r}$ increases.

\subsection{Results for the compact model with $N=2$}

As discussed in the main text, we also performed an exploratory
investigation of the effect of explicit gauge breaking terms using the
compact discretization.  We studied the $N=2$ cAH model, which is the
only one in which a continuous transition is present
\cite{PV-19-AH3d}. In this case we defined the photon mass operator as
\begin{equation}
P_M=- r \sum_{{\bm x},\mu} {\rm Re} \,\lambda_{{\bm x},\mu}\,,
\label{suppl-Pm-compatto}
\end{equation}
and performed simulations using the axial gauge. We set $\lambda_{{\bm x},3}=1$
(this the analogue of the condition $A_{{\bm x},3}=0$ used in the noncompact
case) on all sites. Note that this is possible as we use $C^*$ boundary
conditions also for the compact model.

\begin{figure}[btp]
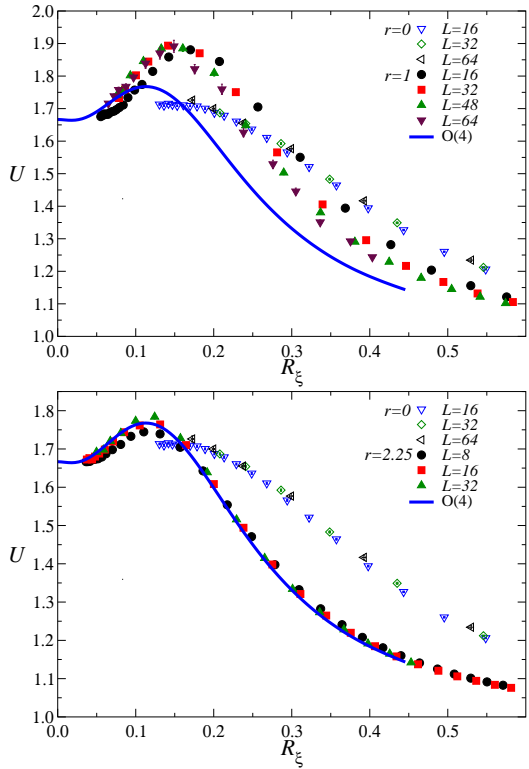

\includegraphics*[width=0.8\columnwidth]{urxi_nccp1_kappa0p0_m1p0.eps}
\includegraphics*[width=0.8\columnwidth]{urxi_nccp1_kappa0p0_m1p5.eps}
\caption{Estimates of $U$ versus $R_\xi$ for the cAH model M1 with
  $N=2$, $g_0^2=\infty$: results for $r=1$ (top panel), and $r=2.25$
  (bottom panel).  The critical couplings are $J_c\approx 0.552$ and
  $J_c\approx 0.328$ for $r=1$ and $r=2.25$ respectively.  We also
  report results for the gauge-invariant model ($r=0$) and the O(4)
  vector model (full line, obtained by large-$L$ extrapolations of MC
  data for the appropriate spin-2 correlations, see
  Ref.~\cite{PV-19-AH3d}).}
\label{suppl-rcritN2}
\end{figure}
As for $N=25$ we studied the behavior of the model with action
$S_{cAH} + P_M$ for two finite values of $r$, $r=1$ and
$r=2.25$.  In Fig.~\ref{suppl-rcritN2} we report the Binder parameter
versus $R_\xi$.  Data for $r=2.25$ are perfectly consistent with an
O(4) behavior, confirming the expected symmetry enlargement.  The
results for $r=1$ instead are still quite far from the O(4) curve,
although they show the correct trend. Except for values of $R_\xi$
where $U$ has a maximum, as $L$ increases from 16 to 64, the data move
towards the O(4) universal curve. Close to the maximum, the behavior
is nonmonotonic, but the data start moving towards the O(4) curve for
$L\ge 48$. 
A thorough investigation of the compact model will be reported 
in a forthcoming publication \cite{BPV-inprep}.


\begin{thebibliography}{99}
\bibitem{footnote-highercha} If one considers the model with compact fields,
a critical behavior associated with the charged AH FP is only observed with 
fields of higher charge $q \ge 2$~\cite{BPV-20-hc}.
\bibitem{MCP} If $y_r < 1/\nu$, Eq.~(\ref{tfrxy}) still holds but one
  has to replace $X$ with the appropriate linear scaling field (see
  Ref.~\cite{Fisher-MCP}), $X = (J - a r - J_c) L^{1/\nu}$. The
  constant $a$ is fixed by the requirement that $J_c(r) = J_c + a r +
  O(r^2)$. Here $J_c(r)$ is the transition value for the model at
  fixed $r$ and $J_c=J_c(0)$. If $y_r < 1/\nu$, the behavior for $J=J_c$
  is controlled by $X = - a r L^{1/\nu}$ and thus, 
  fits of $R$ to functions of $r L^{y_r}$ 
  would give $y_r = 1/\nu$: no information on the relevance or irrelevance of
  the perturbation would be obtained.
\end{thebibliography}
\end{document}